\documentclass[sigconf, nonacm, pbalance]{acmart}

\usepackage{xspace}
\usepackage{amsmath}
\usepackage{algorithm}
\usepackage{algpseudocode}
\usepackage{booktabs}
\usepackage{array}
\usepackage{tabularx}
\usepackage{pifont}
\newcommand{\equalcontributionmark}{\textsuperscript{1,2,\#}}
\newcommand{\correspondingauthormark}{\textsuperscript{1,2,*}}
\begin{document}
\title[RegionFence]{RegionFence: Adaptive Concurrency Control for Blockchain Parallel Transaction Execution}

\author[Jiaying Feng]{Jiaying Feng\equalcontributionmark}
\affiliation[obeypunctuation=true]{\institution{\textsuperscript{1}Institute of Computing Technology, Chinese Academy of Sciences}\\
  \institution{\textsuperscript{2}School of Computer Science and Technology, University of Chinese Academy of Sciences}\\
  \city{Beijing}, \country{China}
}
\email{fengjiaying24s@ict.ac.cn}
\author[Linpeng Jia]{Linpeng Jia\equalcontributionmark}
\affiliation[obeypunctuation=true]{\institution{\textsuperscript{1}Institute of Computing Technology, Chinese Academy of Sciences}\\
  \institution{\textsuperscript{2}School of Computer Science and Technology, University of Chinese Academy of Sciences}\\
  \city{Beijing}, \country{China}
}
\email{jialinpeng@ict.ac.cn}
\author[Yi Sun]{Yi Sun\correspondingauthormark}
\affiliation[obeypunctuation=true]{\institution{\textsuperscript{1}Institute of Computing Technology, Chinese Academy of Sciences}\\
  \institution{\textsuperscript{2}School of Computer Science and Technology, University of Chinese Academy of Sciences}\\
  \city{Beijing}, \country{China}
}
\email{sunyi@ict.ac.cn}

\begin{abstract}
Smart contracts have turned blockchains into general platforms, and their throughput is increasingly limited by transaction execution. Parallel execution can relieve this bottleneck. However, most existing engines apply a single policy within a block, typically optimistic (OCC) or pessimistic (PCC) concurrency control, and thus re-execute or delay whole transactions. Although some database studies have proposed hybrid and adaptive mechanisms, these mechanisms cannot be applied directly to blockchains because the block fixes the serial order, accesses emerge only at run time, and real workloads vary widely.

We propose RegionFence, an adaptive concurrency control mechanism for blockchain parallel transaction execution, based on three key designs. First, we propose regions to refine the granularity of control from whole transactions to the parts that touch contended keys. Second, we design fences to decide at run time how each region handles a conflict. Third, we introduce intra-block and inter-block online learning to anticipate likely conflicts. At the same time, the system still produces the same result as serial execution in consensus order.

We implement RegionFence in go-ethereum and evaluate it on real Ethereum mainnet blocks from different periods of the past five years. Experimental results show that RegionFence achieves an overall speedup of 9.2$\times$ over single-core execution on 64 cores and outperforms OCC by 2.9--5.4$\times$ and PCC by 3.9--9.4$\times$ in speedup across these periods.

\end{abstract}

\maketitle
\hypersetup{pdfauthor={Jiaying Feng, Linpeng Jia, Yi Sun}}
\ifdefined\pdfinfo
\pdfinfo{/Author (Jiaying Feng, Linpeng Jia, Yi Sun)}
\fi
\begin{NoHyper}
\begingroup
\renewcommand\thefootnote{}
\footnotetext{\noindent\textsuperscript{\#} These authors contributed equally to this work.\\
\textsuperscript{*} Corresponding author.}
\endgroup
\end{NoHyper}
\pagestyle{plain}
{\small\noindent\textbf{Code Availability:} Code and artifacts are available at \url{https://github.com/ictbc-feng/RegionFence}.\par}

\section{Introduction}
\label{sec:introduction}

Blockchains have evolved from ledgers of simple payments~\cite{nakamoto2008bitcoin} into general platforms. With smart contracts~\cite{wood2014ethereum}, they now host token exchanges, lending markets, and a growing range of other applications. These platforms order transactions into blocks and then execute them, and each transaction runs contract code that reads and writes keys of a shared state. As consensus protocols have grown faster, the throughput of these blockchains is increasingly limited by transaction execution rather than by ordering~\cite{blockstm,parallelevm,nemo}. Parallel execution is the established remedy for this bottleneck, as multi-core databases have shown at up to thousands of cores~\cite{silo,tictoc,yu2014abyss,2pl-thousand-cores}. However, many blockchain clients still execute the transactions of a block sequentially and leave the available cores unused~\cite{parallelevm,anjana2025conflictspec}. Moving these clients to parallel execution must preserve what sequential execution guarantees.

A blockchain is a replicated system: every node executes the blocks that consensus has ordered, and all nodes must arrive at the same states and receipts. Parallel execution must therefore produce the \emph{serial result}, which consists of the states and receipts of executing the transactions sequentially in block order. This requirement is deterministic serializability with a preset serialization order~\cite{blockstm,parallelevm,spectrum,nemo}.

Existing parallel engines meet this requirement with optimistic or pessimistic concurrency control (OCC or PCC). OCC executes transactions speculatively, validates them in block order, and re-executes those that read stale values~\cite{blockstm,parallelevm,nemo}. PCC instead serializes conflicting transactions, usually based on access sets known in advance~\cite{solana,sui-lutris,parblockchain}. A conflict therefore incurs wasted work under OCC and idle cores under PCC.

\suppressfloats[t]
\begin{figure}[htbp]
\centering
\includegraphics[width=\columnwidth]{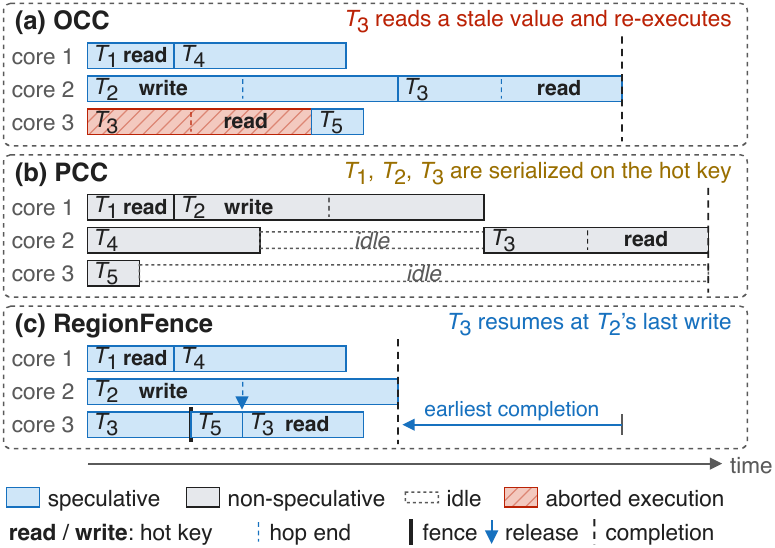}
\caption{Example schedules of one block on three cores. $T_3$ depends on $T_2$ only through the hot key $k_P$: (a)~OCC re-executes $T_3$, (b)~PCC delays it, and (c)~RegionFence resumes it at $T_2$'s last write.}
\Description{Three stacked panels show the same five-transaction block on three cores under OCC, PCC, and RegionFence; each row is a core and each bar a transaction. Under OCC, T3's first run is discarded; under PCC, T1, T2, and T3 run in turn while cores idle; under RegionFence, T3 stops at a fence, T5 runs on its core, and T3 resumes when T2 releases the hot key, so the block finishes first, then under OCC, then under PCC.}
\label{fig:motivation}
\end{figure}

Figure~\ref{fig:motivation} illustrates both costs with an example. Consider a block that contains transactions $T_1$ to $T_5$ of a decentralized exchange, where a swap trades tokens by updating the reserves of a pool. $T_1$ is a price-feed update that reads key $k_P$, which stores the reserves of a popular pool P. $T_2$ and $T_3$ are swaps routed through two pools. $T_2$ trades in P and then in a quiet pool, while $T_3$ trades in another quiet pool and then in P, so $T_3$ reads the value of $k_P$ that $T_2$ writes. $T_4$ and $T_5$ are unrelated token transfers.

OCC runs $T_2$ and $T_3$ concurrently without waiting, so $T_3$ reads $k_P$ before $T_2$ has written it. When $T_3$ is validated, $k_P$ has changed since its read, so the engine aborts $T_3$ and re-executes it from the start (Figure~\ref{fig:motivation}(a)). PCC avoids this abort, but $T_1$, $T_2$, and $T_3$ all access $k_P$, so they run one after another while other cores remain idle (Figure~\ref{fig:motivation}(b)).

In this example, both costs are unnecessary because $T_3$ needs only the value that $T_2$ writes to $k_P$. $T_3$ could execute its first hop concurrently with $T_2$, wait at its read of $k_P$, and resume once $T_2$ completes its last write, even though $T_2$ has not finished (Figure~\ref{fig:motivation}(c)). Neither family produces this schedule because most engines apply one policy to a whole block~\cite{nemo}.

Recent blockchain studies reduce the cost of a conflict by re-executing only the operations or subtransactions it affects~\cite{parallelevm,loom,spectrum}. Others shorten execution with hardware acceleration~\cite{mtpu,mpt-gpu}, and permissioned systems gain flexibility from reordering transactions or switching architectures~\cite{fabric,fabricpp,adachain}. These approaches lower the cost of conflicts and of execution, but how a public blockchain should handle each conflict during execution remains largely open.

Database research offers a starting point for this question, as it has explored more flexible concurrency control. Some studies combine multiple policies at finer granularities~\cite{mocc,sundial,cormcc,tebaldi,hdcc,aocc,polyjuice}, some adapt to contention or the workload at run time~\cite{rebirth-retire,caracal,txnsails}, and others make writes visible before transactions finish~\cite{bamboo,pwv,ic3}. However, these designs are difficult to adopt directly because blockchain execution differs from database transaction processing in three respects. First, the block fixes the serial order before execution, which leaves little room for techniques that choose the order at run time~\cite{sundial,aria}. Second, accesses emerge only at run time, so policies built on known access sets~\cite{calvin,bohm}, transaction types~\cite{tebaldi}, or offline training~\cite{polyjuice} become less effective. Third, the actual workloads of mainstream blockchains are complex within a transaction, within a block, and across blocks, so a policy tuned for one case rarely fits another (Section~\ref{sec:observations}).

These differences call for adaptive concurrency control designed for blockchain parallel transaction execution. Such a mechanism should produce such schedules on real workloads, and building it raises three challenges.

\textbf{Challenge 1: granularity of control.} A conflict involves only part of a transaction, so controlling an entire transaction delays operations that conflict with no other transaction. Conversely, controlling every access imposes overhead on the many accesses that never conflict.

\textbf{Challenge 2: selection and composition of policies.} A conflict can be resolved in several ways, and the most effective choice depends on costs that vary with the workload and the number of cores. Moreover, transactions under different policies coexist within a block and may read uncommitted values, yet they must remain free of deadlocks, limit cascading aborts, and still produce the serial result.

\textbf{Challenge 3: online prediction of conflicts.} A wait is beneficial only if it precedes the conflicting access, so the engine must anticipate which transactions will conflict. Such knowledge is rarely available in advance for open and evolving contract workloads, and it must therefore be acquired during block execution at low overhead. The same applies to how many blocks can usefully execute together.

To address these challenges, we control a transaction only where it touches a contended key, decide each wait from costs measured online, and learn conflicts from execution itself. We realize these ideas in RegionFence, an adaptive concurrency control mechanism for blockchain parallel transaction execution. This paper makes the following contributions:
\begin{itemize}
 \item \textbf{Regions.} We introduce the \emph{region} as the unit of concurrency control. A region is the part of a transaction between its first read of a contended key and its last write to it. Concurrency control acts only at the two ends, so accesses that never conflict need no coordination during execution (Section~\ref{sec:regions}).
 \item \textbf{Fences.} We design \emph{fences} that decide at the start of a region whether a read proceeds or waits for predicted writers. At the end of a region, a writer exposes its value early to waiting readers if the write is likely final, and an aborted writer marks the keys it wrote by how likely their values are to change. We prove that these decisions preserve the serial result and guarantee termination (Section~\ref{sec:fences}).
 \item \textbf{Intra-block and inter-block online learning.} We design \emph{conflict profiles} that predict the likely writers of a contended key from features known before a transaction runs, such as its contract, function, and address arguments. They learn from the read and write sets that execution reveals, without declared access sets, static analysis, or offline training. Some contended keys are new to a block and others recur across blocks, so learning runs at two time scales. Intra-block learning reacts to conflicts as soon as they appear, and inter-block learning accumulates what recurs and sizes a window of blocks that execute together (Section~\ref{sec:learning}).
 \item \textbf{Implementation and evaluation.} We implement RegionFence in go-ethereum, along with the OCC (Block-STM) and PCC (runtime locking) baselines (Section~\ref{sec:implementation}). We design synthetic workloads for blockchains rather than reuse database benchmarks such as TPC-C~\cite{tpcc} and SmallBank~\cite{smallbank}, and we replay real Ethereum mainnet blocks from different periods of the past five years (Section~\ref{sec:evaluation}). Experimental results show that RegionFence achieves an overall speedup of 9.2$\times$ over single-core execution on 64 cores and outperforms OCC by 2.9--5.4$\times$ and PCC by 3.9--9.4$\times$ in speedup across these periods.
\end{itemize}

\section{Background and Observations}
\label{sec:background-and-observations}

This section reviews parallel transaction execution on blockchains and existing approaches (Section~\ref{sec:background}), and then characterizes mainnet workloads to derive design implications (Section~\ref{sec:observations}).

\subsection{Parallel Execution on Blockchains}
\label{sec:background}

\begin{table*}[t]
\caption{Representative parallel execution systems. \emph{Control unit}: the scope of one decision.}
\label{tab:design-space}
\footnotesize\setlength{\tabcolsep}{5pt}\renewcommand{\arraystretch}{1.0}
\begin{tabularx}{\textwidth}{@{}l>{\raggedright\arraybackslash}p{3.6cm}>{\raggedright\arraybackslash}X>{\raggedright\arraybackslash}p{3.2cm}>{\raggedright\arraybackslash}p{4.0cm}@{}}
\toprule
System & Control unit & Conflict information & Decision time & Wait condition\\
\midrule
Block-STM~\cite{blockstm} & Transaction & Writes of aborted incarnations & Upon abort & Writer re-execution\\
ParallelEVM~\cite{parallelevm} & Operation & None & Validation phase & None\\
Sui Lutris~\cite{sui-lutris} & Object version & Declared objects & Before execution & Completion of earlier versions\\
Spectrum~\cite{spectrum} & Hot-key read & Partial pre-acquired sets & Before execution & Final commit of writers\\
OCToPus~\cite{octopus} & Atomic operation & Transfer semantics & Upon insufficient funds & Serial fallback\\
Crystality~\cite{crystality} & Function fragment & Programmer-defined scopes & Development time & None\\
Fabric++~\cite{fabricpp} & Transaction & Simulated read and write sets & Ordering phase & None\\
NeuChain~\cite{neuchain} & Transaction & Reserve table & Epoch end & None\\
Loom~\cite{loom} & Subtransaction & Block pre-execution & After pre-execution & Conflicting subtransactions\\
AdaChain~\cite{adachain} & Entire system & Workload features, throughput & Per episode & Determined by architecture\\
\midrule
\textbf{RegionFence} & \textbf{Region on a contended key} & \textbf{Learned from execution} & \textbf{Run time, at region start} & \textbf{Writer's last write}\\
\bottomrule
\end{tabularx}
\end{table*}

A blockchain is a replicated state machine maintained by mutually distrusting nodes. Its state maps \emph{keys}, namely account balances, nonces, code, and contract storage slots, to values. Consensus orders transactions into blocks, and every node executes them in block order. Each transaction invokes contract code, so its read and write sets are determined at run time and are difficult to predict before execution~\cite{spectrum}. Because all nodes must reach identical states and receipts, a block defines a preset serialization order, and any concurrent execution must produce the serial result~\cite{blockstm}. Sequential execution satisfies this requirement but cannot exploit multiple cores, which motivates parallel execution.

A parallel execution engine runs transactions concurrently and may execute a transaction multiple times, where each execution is an \emph{incarnation}~\cite{blockstm}. Incarnations \emph{publish} their writes as versions in a multi-version store, and transactions commit in block order.

Two transactions conflict if they access the same key and at least one of the accesses is a write. Under a preset serialization order, only one of the three conflict types constrains execution. For $i<j$, a write-after-write (WAW) conflict requires no coordination, because each transaction writes its own version and in-order commit makes the version of $T_j$ the later one. A write-after-read (WAR) conflict, where $T_i$ reads and $T_j$ writes, also requires no coordination, because $T_i$ reads only versions below $i$. Only a read-after-write (RAW) conflict, where $T_j$ reads what $T_i$ writes, requires coordination, because $T_j$ must observe the version of $T_i$.

We say that $T_j$ \emph{depends} on $T_i$, written $T_i \to T_j$, if $T_j$ reads a key whose latest preceding writer is $T_i$. Because dependencies follow block order, they cannot form cycles, unlike those in a database that determines its own serialization order. Concurrency control therefore reduces to one question: how a reader should proceed when an earlier writer of the same key has not finished. We call such a key \emph{contended}.

Optimistic concurrency control (OCC)~\cite{kung1981occ} allows the reader to proceed and corrects the outcome afterwards. Transactions execute speculatively on the latest visible versions, validation in block order detects stale reads, and the affected transactions re-execute. In Block-STM~\cite{blockstm}, the writes of an aborted incarnation are marked as estimates, and a reader that encounters an estimate aborts and restarts after the writer's next incarnation finishes. Later systems reduce the cost of repair: ParallelEVM~\cite{parallelevm} re-executes only the affected operations, Spectrum~\cite{spectrum} rolls back to the conflicting read, and Reddio~\cite{reddio} defers conflicting transactions to a later batch. In all of these systems, a conflict is detected only after speculative work has been performed.

Pessimistic concurrency control (PCC) instead blocks the reader by ordering conflicting transactions in advance. Following deterministic concurrency control~\cite{calvin,bohm}, Sealevel~\cite{solana} and Sui Lutris~\cite{sui-lutris} schedule transactions from declared accounts or objects, ParBlockchain~\cite{parblockchain} from known read and write sets, dBTM~\cite{anjana2025conflictspec} from conflict specifications, and MTPU~\cite{mtpu} from a dependency graph, and other designs let validators replay a miner's schedule~\cite{dickerson2017concurrency,anjana2019occ}. Runtime locking, in contrast, acquires locks at each access and releases them at commit in block order. These designs serialize conflicting transactions in their entirety, which delays non-conflicting operations, and schedules fixed in advance require access information that smart contracts rarely expose.

Between these two extremes, some systems use partial information to decide which reads wait. Spectrum blocks reads of hot keys until writers in pre-acquired sets reach final commit, NEMO~\cite{nemo} extends Block-STM with static hints and priorities, and DMVCC~\cite{dmvcc} exposes writes once program analysis shows that no abortable statement remains. This information must still be obtained before execution, and apart from DMVCC, readers are held until the writer commits or passes validation.

Other systems prevent such conflicts from arising. Permissioned systems and block builders reorder or recompose blocks~\cite{loom,neuchain,gbdcc,fabric,fabricpp,fafo,anthemius}, which validators of an already ordered block cannot do. Semantic approaches eliminate specific classes of conflicts~\cite{rapidlane,octopus,crystality}, often by rewriting contracts, but other read-modify-write conflicts remain. AdaChain~\cite{adachain} learns a system-wide architecture for permissioned blockchains, a decision much coarser than a single conflict.

Table~\ref{tab:design-space} compares representative systems, which leave three limitations. Most systems decide per block, transaction, or architecture rather than per conflict. Readers typically wait until a writer finishes or commits, even when its last write to the key occurs much earlier. Waits also rely on access information supplied before execution rather than learned from it.

\begin{figure*}[t]
\centering
\includegraphics[width=\textwidth]{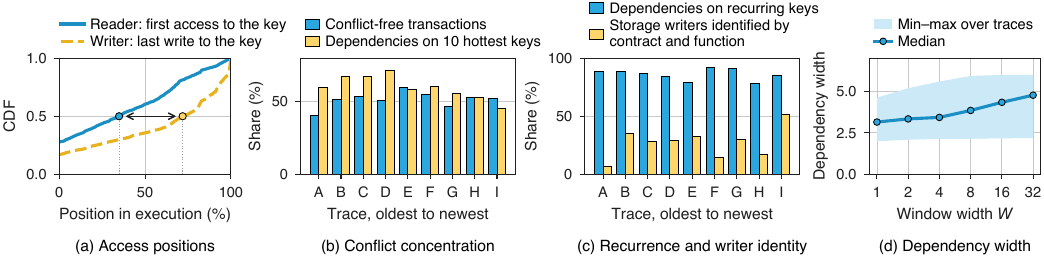}
\caption{Workload measurements on the traces of Table~\ref{tab:traces}: (a)~positions of the reader's first access and the writer's last write, (b)~conflict-free transactions and dependencies on hot keys, (c)~recurring keys and writers identified by contract and function, and (d)~dependency width as $W$ grows. Panel (c) uses blocks 50--99 for recurrence and blocks 20--99 for writers, with 128 transactions in flight, and (d) includes every account load.}
\label{fig:measurements}
\Description{Four panels characterize mainnet conflicts: positions of readers' first accesses and writers' last writes, conflict-free transactions and hot-key dependencies, recurrence of keys and attribution of writers, and dependency width as the number of concurrent blocks increases.}
\end{figure*}

\subsection{Observations on Mainnet Workloads}
\label{sec:observations}

Whether these limitations matter in practice depends on real-world workloads, which prior studies show to be complex~\cite{loom,adachain,mtpu}. We therefore characterize how conflicts arise within a transaction, within a block, and across blocks. We replay Ethereum mainnet blocks on their recorded pre-states, derive dependencies from the recorded read and write sets, and locate each access by the fraction of instructions executed before it. Table~\ref{tab:traces} lists the nine \emph{traces} used in both the measurements and the evaluation. We consider accesses to storage slots, account balances, and nonces, excluding those of the beneficiary, and only writes that change a value.

\begin{table}[t]
\caption{Mainnet traces of 100 consecutive blocks each, oldest to newest.}
\label{tab:traces}
\footnotesize
\renewcommand{\arraystretch}{0.95}
\begin{tabular*}{\columnwidth}{@{\extracolsep{\fill}}clcr@{}}
\toprule
Trace & First block & Date & Txs \\
\midrule
A & 12,499,900 & 2021-05-24 & 20,171 \\
B & 13,809,900 & 2021-12-15 & 17,167 \\
C & 14,500,000 & 2022-04-01 & 16,810 \\
D & 14,999,900 & 2022-06-21 & 18,408 \\
E & 16,619,900 & 2023-02-13 & 15,308 \\
F & 17,873,600 & 2023-08-08 & 14,968 \\
G & 20,079,900 & 2024-06-13 & 17,011 \\
H & 22,449,900 & 2025-05-10 & 18,232 \\
I & 26,060,700 & 2026-09-26 & 23,260 \\
\bottomrule
\end{tabular*}
\end{table}

\textbf{Observation 1: conflicts cover only part of a transaction.} A dependency constrains only a span of the reader's and the writer's execution, and their other operations are unaffected by it. For example, the reader's first access to the key occurs at a median of 35\% of its execution and the writer's last write at 72\% (Figure~\ref{fig:measurements}(a)). Transaction-level control therefore delays unrelated operations, so control must be finer-grained.

\textbf{Observation 2: conflicts are concentrated on a few keys.} Within a block, many independent transactions coexist with a few hot keys that attract most conflicts. For example, about half of a block's transactions conflict with no other transaction, and the ten hottest keys account for 46\%--72\% of its dependencies (Figure~\ref{fig:measurements}(b)). A single policy suits only one of these two groups, so speculation and blocking must coexist within a block.

\textbf{Observation 3: conflicts recur but are hard to attribute.} Contended keys recur across blocks, which makes conflicts learnable, but new keys continue to emerge and coarse-grained information does not identify their writers. For example, 79\%--92\% of a block's dependencies involve keys that carried a dependency in the previous 50 blocks, whereas the contract and function selector identify only 7\%--51\% of the active storage writers at a read (Figure~\ref{fig:measurements}(c)). Writers should therefore be predicted from finer features, learned online within and across blocks.

\textbf{Observation 4: parallelism lies across blocks.} A single block offers limited parallelism, and executing blocks concurrently exposes more of it only if reads wait for the correct writers. For example, the \emph{dependency width} of a trace, its total execution time divided by the summed critical-path lengths of its blocks, has a median of 3.15 and reaches only 4.77 with $W=32$ concurrent blocks, whereas the region-level bound reaches 12.26 (Figure~\ref{fig:measurements}(d)). Execution should therefore span blocks, and the number of concurrent blocks must be determined online.

\section{RegionFence Overview}
\label{sec:overview}

RegionFence is a conflict-level concurrency control mechanism: it takes each conflict, rather than a transaction or a block, as the unit of decision. It replaces sequential execution on a node: it executes the transactions of blocks ordered by consensus in parallel and produces the same state and receipts, so consensus, contracts, and the state format remain unchanged. Because a node replays blocks that are already ordered, RegionFence cannot reorder or drop transactions, and all of its control lies in the timing of reads.

\textbf{Core Abstractions.} RegionFence builds a schedule from three abstractions: regions, fences, and releases. A \emph{region} defines what is controlled: the span of a transaction that a dependency constrains, a unit of control between an operation and a transaction. A \emph{fence} defines how it is controlled: an ordering constraint, established at run time, between a reader and a predicted writer. A \emph{release} defines when control ends: the constraint dissolves at the writer's last write to the key rather than at its commit. A schedule is thus the result of ordering constraints that are established and removed during execution. Certification at commit is separate from these constraints: it checks every transaction in block order, so any combination of constraints is safe, and wrong estimates cannot affect the result.

\begin{figure}[t]
\centering
\includegraphics[width=\columnwidth]{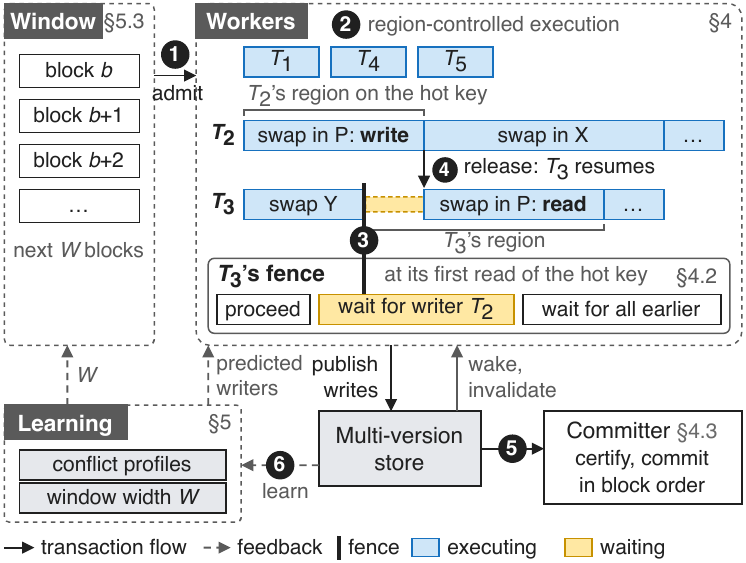}
\caption{RegionFence workflow on the running example. Numbered steps follow one block, in which $T_3$ waits at its first read of the hot key $k_P$ until $T_2$'s last write to it.}
\label{fig:overview}
\Description{A workflow diagram shows the execution window, workers, multi-version store, committer, and learning module. Six numbered steps follow transaction T3 as it waits on the hot key, resumes after T2 releases the value, commits after certification, and supplies learning feedback.}
\end{figure}

\textbf{System Architecture.} RegionFence consists of five components, shown in Figure~\ref{fig:overview}. The \emph{execution window} admits the next $W$ blocks for concurrent execution. The \emph{workers} execute transactions under region control and suspend a transaction that must wait, so that a blocked transaction does not occupy a core. They serve the lowest-indexed ready transaction first, so the commit frontier always progresses. The \emph{multi-version store} holds the versions of each key together with their registered readers. Blocks in the window form one transaction sequence, so a transaction may read a version written by an uncommitted transaction of an earlier block. The \emph{committer} certifies transactions in block order and commits them. The \emph{learning module} maintains \emph{conflict profiles}, which estimate the likely writers of each contended key, and the window width.

The components interact through a control path, a data path, and a feedback path. On the control path, a transaction consults the conflict profile of a key at a fence and decides whether to proceed or to wait. Writers and aborted transactions make the corresponding decisions, namely whether to release a value and whether their versions stay visible. On the data path, transactions publish their writes to the store, which wakes waiting readers and invalidates registered readers of changed versions. On the feedback path, finished incarnations and measured costs flow to the learning module.

\textbf{Transaction Lifecycle.} The lifecycle of a transaction consists of six steps, illustrated here on the block of Figure~\ref{fig:motivation}. \ding{172}~The execution window admits the next $W$ blocks. \ding{173}~Workers execute the transactions: $T_1$, $T_4$, and $T_5$ proceed without waiting, and $T_3$ completes its swap through pool Y. \ding{174}~When $T_3$ first reads $k_P$, it enters a region, and its fence consults the conflict profile of $k_P$. The profile identifies $T_2$ as a likely writer, so the fence orders $T_3$ after $T_2$, and $T_3$ suspends and frees its worker, which then executes $T_5$.

The remaining steps release the reader, commit the transactions, and return evidence to the learning module. \ding{175}~Because $T_3$ is waiting, $T_2$ releases $k_P$ immediately after its last write to it. $T_3$ resumes while $T_2$ is still executing its swap through pool X, and $T_2$ publishes its remaining writes when it finishes. \ding{176}~The committer certifies each transaction in block order and commits it, and an incarnation that fails certification re-executes. \ding{177}~Finished incarnations reveal which keys they wrote, invalidations reveal which keys are contended, and waits and re-executions reveal what they cost. RegionFence feeds this evidence into the conflict profiles, its cost estimates, and the window width, so that later decisions follow the workload.

\textbf{Handling Mispredictions.} RegionFence handles three kinds of constraint error. A superfluous constraint, on a predicted writer that does not write the key, delays the reader until that writer finishes and lowers the estimate behind it. A constraint removed too early, by the release of a value that later changes, invalidates the readers of that value, which re-execute. A missing constraint, on a writer that was not predicted, is detected when the writer publishes, which invalidates the registered readers and adds the key to the contended set. These errors therefore affect performance but not correctness.

\section{Regions and Fences}
\label{sec:regions-fences}

This section presents regions, fences, and releases, the mechanisms that control a conflict, and proves their correctness. Regions address Challenge~1 (Section~\ref{sec:regions}), and fences and their composition address Challenge~2 (Sections~\ref{sec:fences} and~\ref{sec:correctness}).

\subsection{Region Model}
\label{sec:regions}

\textbf{Region Definition.} A region is the part of a transaction that a dependency constrains. We model a transaction as a sequence of observations and writes on a key-value state, where an \emph{observation} is an access whose outcome depends on the value of a key. Observations determine a transaction's writes and result, so a schedule produces the serial result if every observation returns what it returns in serial execution~\cite{bernstein1987concurrency}. A dependency $T_i \to T_j$ through key $k$ therefore constrains exactly one pair of points: $T_j$ must observe $k$ after the last write of $T_i$ to it. All other parts of the two transactions may overlap.

RegionFence confines concurrency control to the two ends of each region. A transaction has one \emph{region} for each contended key it observes, which extends from its first observation of the key to its last write to the key, or to its end if it never writes the key. Whereas two-phase locking holds a key until commit, a region ends at the last write. A transaction makes decisions only at the ends of its regions and is unconstrained outside them.

\textbf{Contended Key Detection.} RegionFence infers contended keys from conflicts rather than from access frequency. A frequently read key need not be contended, because its reads may rarely precede an earlier writer's publication. A key joins the set $K_c$ of contended keys the first time an earlier writer invalidates a read of it. The other keys of the same object that the writer changed join as well, because a contended object rarely occupies one key.

Only reads of keys in $K_c$ pass through a fence, with one exception for re-executed transactions. The failure of such a transaction is direct evidence: its read of a key on which an earlier incarnation failed, or of a version from an uncommitted writer, also enters a region. Accesses that never conflict therefore incur no coordination.

\textbf{Region Entry.} A region starts at a transaction's first observation of a key. An access is an observation only if the transaction's outcome depends on the value it returns, so many conflicts between reads and writes impose no constraint.

RegionFence excludes three classes of accesses from regions through semantic concurrency control. Independent fields of an object are separate keys, so an access to one does not conflict with a write to another. An update to a value that the transaction does not read is a commutative update, applied in block order~\cite{rapidlane,octopus}. A predicate read observes only its outcome, which is re-evaluated at commit, so it conflicts only if the outcome changes~\cite{rapidlane}. Section~\ref{sec:implementation} maps these classes to the EVM.

\textbf{Region Exit.} A region ends at the writer's last write, where a \emph{release} exposes the value to readers. The value is then final for the writer's incarnation, so a reader need not wait for the writer to finish. A release is a form of early write visibility, as in early lock release and Bamboo~\cite{bamboo}. However, a writer cannot determine whether a write is its last, and its incarnation may still abort. The end of a region is therefore a prediction, which Section~\ref{sec:fences} treats as a decision.

\subsection{Fence Protocol}
\label{sec:fences}

\textbf{Cost Model.} RegionFence makes every decision at a region with one cost-based rule. Each decision determines whether a transaction should read an uncommitted version or wait for it. OCC and PCC fix this choice for a whole block, whereas RegionFence makes it for each conflict.

The rule applies at three points in the life of a transaction. At the start of a region, an earlier writer may not have written the key yet, so the reader can proceed on the visible version or wait for that writer. At the end of a region, the writer may write the key again or abort, so it can release the value or hold it until it publishes. After an abort, the next incarnation may change the versions of the aborted one, so readers can proceed on them or wait for that incarnation. At each point, one choice risks a re-execution and the other a delay.

The rule compares the expected costs of the two possible errors. Consider an event of estimated probability $\pi$, such as a predicted writer writing the key. Acting on the event costs $A$ if it does not occur, and not acting costs $B$ if it does. Comparing the expected costs $(1-\pi)A$ and $\pi B$, acting is beneficial when
\begin{equation}\label{eq:rule}
\pi \;\ge\; \tau = \frac{A}{A + B}.
\end{equation}

The rule combines a probability from learning with a threshold from run-time measurements. The probability $\pi$ depends on the key (Section~\ref{sec:learning}), whereas the threshold $\tau$ depends on the machine and the workload, so RegionFence measures $A$ and $B$ in time during execution. Delays are weighted by $u$, the idle share of worker time: while other work occupies the cores a delay costs little, and while they idle it costs nearly its duration. A needless invalidation is weighted by $1-u$ instead, because a re-execution consumes processors only while they are busy. Because $\tau$ follows these measurements, the choice adapts to the workload and the number of cores.

\begin{algorithm}[t]
\caption{Fence: a reader at the start of a region}
\label{alg:fence}
\small
\algtext*{EndIf}\algtext*{EndFor}\algtext*{EndFunction}
\begin{algorithmic}[1]
\Function{Fence}{$T_j, k$} \Comment{$T_j$ starts a region on $k$}
  \If{$k$ is globally written}
    \State wait until $T_j = T_{\min}$ \label{line:front}
  \Else
    \State $C \gets$ the nearest predicted writers of $k$ below $j$ \label{line:candidates} \Comment{\S\ref{sec:intra-block}}
    \For{$T_i \in C$, nearest first}
      \If{$T_i$ is unfinished and $\hat p_k(T_i) \ge p^\ast$} \label{line:pstar}
        \State yield until $T_i$ releases or publishes $k$, or finishes
        \State \textbf{if} $T_i$ wrote $k$ \textbf{then break} \label{line:stop}
        \State lower the feature that predicted $T_i$ \Comment{unnecessary wait} \label{line:vain}
      \EndIf
    \EndFor
  \EndIf
  \State $v \gets \mathit{vis}_j(k)$ \Comment{waits if the version is labelled}
  \State \textbf{if} some $T_i$, $i<j$, is unfinished \textbf{then} register $T_j$ on $v$
  \State \Return $v$
\EndFunction
\end{algorithmic}
\end{algorithm}

\textbf{Read Fences.} A \emph{fence} decides whether a transaction entering a region proceeds or waits (Algorithm~\ref{alg:fence}). In the algorithm, $T_{\min}$ is the \emph{commit frontier}, the lowest uncommitted transaction, and $\mathit{vis}_j(k)$ is the value that a read of $k$ by $T_j$ returns: the latest version published below $j$ in the block of $T_j$, else the value set by the state changes $\Gamma_b$ applied at the start of that block $b$, else the latest earlier version.

The fence decides whom to wait for and whether waiting is beneficial. For the first part, it asks the conflict profile of $k$ for \emph{predicted writers}, namely unfinished earlier transactions that are likely to write $k$ (Section~\ref{sec:intra-block}), each with an estimate $\hat p_k$ (line~\ref{line:candidates}). For the second part, Equation~\eqref{eq:rule} compares two errors. Waiting for a transaction that does not write $k$ loses $c_w$, and proceeding past one that does loses $c_r$ to a re-execution. Thus $T_j$ waits for a predicted writer if $\hat p_k \ge p^\ast$, where $p^\ast = u c_w / (u c_w + c_r)$ (line~\ref{line:pstar}).

A fence examines predicted writers in order of proximity. $T_j$ visits them from nearest to farthest, because the nearest writer's value is the one it will read. It stops at the first that wrote $k$, and an unnecessary wait lowers the estimate of the feature that predicted the writer within this fence (lines~\ref{line:stop}--\ref{line:vain}). If no wait is warranted, $T_j$ proceeds: it reads the visible version and, while some earlier transaction is unfinished, registers as a reader of that version, so that a later change invalidates it. A \emph{globally written} key, which nearly every earlier transaction writes, rarely has a stable version before the frontier, so a read of it waits until $T_j$ reaches the commit frontier (line~\ref{line:front}).

\begin{algorithm}[t]
\caption{Release and recovery: a writer}
\label{alg:writer}
\small
\algtext*{EndIf}\algtext*{EndFor}\algtext*{EndFunction}
\begin{algorithmic}[1]
\Function{OnWrite}{$T_i, k, m$} \Comment{$m$-th write of $k \in K_c$}
  \If{the write changes $k$ and $\hat q_k(m) \ge \theta$}
    \If{a reader waits for $T_i$} \label{line:release}
      \State release $k$: publish its value, wake the readers
    \Else
      \State hold it; release it once a reader waits \label{line:hold}
    \EndIf
  \EndIf
\EndFunction
\Function{OnAbort}{$T_i$}
  \For{each key $k$ that $T_i$ wrote}
    \If{$k$ failed or rewrites, and $\hat r_k$ reaches its threshold} \label{line:label}
      \State label $k$ failed or rewrite: readers wait
      \State \textbf{if} $\hat r_k$ is high enough \textbf{then} invalidate its readers \label{line:cascade}
    \Else
      \State leave $k$ tentative: readers proceed \label{line:tentative}
    \EndIf
  \EndFor
\EndFunction
\end{algorithmic}
\end{algorithm}

\textbf{Early Release.} A writer applies the same rule to decide whether to release a value. When $T_i$ writes $k$ for the $m$-th time, it does not know whether the write is final. A release fails if $T_i$ writes $k$ again or its incarnation aborts, and in either case the readers re-execute. Let $G$ be the time a release saves waiting readers if the value is final, $L$ the time they lose if it is not, and $o$ the writer's own cost. With $\hat q_k(m)$ the estimated probability that this write is the last, and $\eta$ the fraction of releases whose incarnation survives, Equation~\eqref{eq:rule} becomes
\begin{equation}\label{eq:release}
\hat q_k(m) \;\ge\; \theta = \frac{L + o}{(L + G)\,\eta}.
\end{equation}

RegionFence performs releases on demand, because a release has value only for readers that need it (Algorithm~\ref{alg:writer}, lines~\ref{line:release}--\ref{line:hold}). The writer releases immediately if a reader waits. Otherwise, it holds the value, sends it at a later access once a reader waits, and discards it if it writes the key again.

\textbf{Cascading Abort Control.} RegionFence limits cascading aborts with version states. Releases expose uncommitted values, so schedules are recoverable but do not avoid cascading aborts~\cite{bernstein1987concurrency}. An invalidated incarnation re-executes from the start, but the versions it wrote remain. Hiding them blocks later readers unnecessarily if the values do not change, whereas exposing them propagates the abort if they do. When $T_i$ aborts, it therefore assigns each key it wrote one of three version states (lines~\ref{line:label}--\ref{line:tentative}). Readers wait for the next incarnation on a \emph{failed} key, the written key on which a publication invalidated the incarnation, and on a \emph{rewrite} key, whose versions from aborted incarnations usually differ from the committed ones. Any other key is \emph{tentative}: its version stays visible and readers proceed on it, so they remain valid if the next incarnation writes the same value.

Two further mechanisms handle readers that have already read such a version. Without intervention, earlier readers of such a version would abort one at a time along a dependency chain, so an abort invalidates them immediately when the key is likely to change (line~\ref{line:cascade}). A key that most transactions read and write forms a chain in which a finished writer may depend on a stale read. Before observing such a key, $T_j$ follows the writers' reads of it, invalidates a writer whose read has changed, and waits for it to finish.

\subsection{Correctness}
\label{sec:correctness}

RegionFence mixes multiple policies within one block: one read may wait for a predicted writer, another may read a released value, and a third may proceed on the version of an unfinished transaction. We show that every such mixture produces the serial result and terminates.

We assume that a transaction is deterministic in its observations, so that they determine its writes and receipt, and that every execution terminates, even on an inconsistent state. Let $T_1, T_2, \ldots$ be the transactions of consecutive blocks, and let $\sigma_j$ and $\rho_j$ be the state and receipt after $T_j$ in serial execution. Correctness is independent of the policy applied to each conflict, because RegionFence separates scheduling from certification. Every transaction is subject to one certification rule~\cite{blockstm,spectrum,seer}: \emph{a transaction may commit only if every observation it made equals the same observation on the committed state at its commit point}. Forward validation keeps registered reads current, and backward validation checks every other observation against the committed state. Because commits follow block order, reads of released values remain recoverable, and execution needs no consistent state (no opacity).

\textbf{Theorem 1 (Serial result).} \emph{The committed states and receipts are the serial result.}

\emph{Proof sketch.} The proof is by induction on $j$. Once $T_1, \ldots, T_{j-1}$ have committed with state $\sigma_{j-1}$, every version below $j$ is committed and no longer changes. The certification rule requires the observations of $T_j$ to equal those on $\sigma_{j-1}$, so by determinism $T_j$ produces the writes and receipt of serial execution, which yields $\sigma_j$ and $\rho_j$. $\square$

\textbf{Theorem 2 (Progress).} \emph{The waits-for graph is acyclic, and every transaction commits.}

\emph{Proof sketch.} Every wait targets a lower index or the commit frontier, so the waits-for graph is acyclic. The frontier transaction $T_{\min}$ therefore cannot wait indefinitely, and it is scheduled first. Because it observes only committed values, which are stable, it passes certification, and the frontier advances. $\square$

\textbf{Corollary 1 (Policy composition).} \emph{Theorems~1 and~2 hold for every mixture of policies within a block, that is, for every contended set, fence action, threshold, release, version state, estimate, and window width $W \ge 1$, so nodes whose estimates differ commit the same result.} The proofs rely only on the certification of every observation and on waits that target lower indices, and they make no assumption about the policy applied to a conflict.

\section{Intra-Block and Inter-Block Learning}
\label{sec:learning}

RegionFence estimates three probabilities during execution to address Challenge~3, the online prediction of conflicts. At a fence, the reader of a dependency is known, whereas most writers have not yet accessed the key. Let $\hat p_k(\phi)$ denote the probability that a transaction with feature $\phi$ writes key $k$, $\hat q_k(m)$ the probability that the $m$-th write of $k$ by an execution is its last, and $\hat r_k$ the probability that a version left at $k$ by an aborted execution changes.

These estimates constitute an online estimation problem with full-information feedback. A completed transaction reveals its write set regardless of whether a reader waited for it, so the feedback is independent of the decision. RegionFence therefore requires no exploration, in contrast to the bandit-based tuning of AdaChain~\cite{adachain}.

\begin{figure}[t]
\centering
\includegraphics[width=\columnwidth]{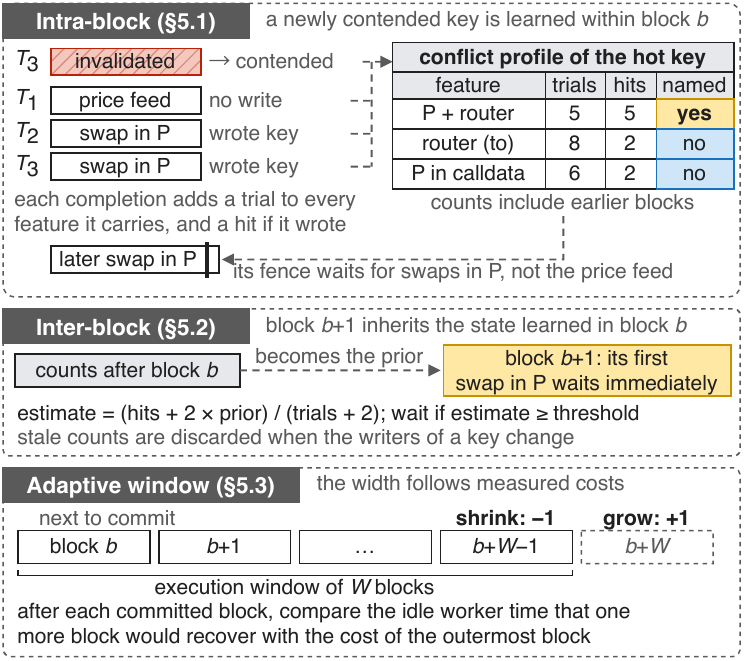}
\caption{Intra-block learning (top), inter-block learning (middle), and window adaptation (bottom) on the running example.}
\label{fig:learning}
\Description{Three stages illustrate learning on the running example: an invalidation identifies a contended key within a block, writer evidence carries across blocks, and measured execution costs adjust the number of blocks in the window.}
\end{figure}

RegionFence learns at two time scales, because some contended keys are new and others recur. Intra-block learning captures new conflicts (Section~\ref{sec:intra-block}), and inter-block learning accumulates recurring ones (Section~\ref{sec:inter-block}). The same feedback determines the number of concurrent blocks (Section~\ref{sec:window}). Figure~\ref{fig:learning} illustrates the three mechanisms on the running example.

\subsection{Intra-Block Learning}
\label{sec:intra-block}

\textbf{Conflict Profiles.} RegionFence predicts writers from features that are available before execution and finer-grained than the invoked code. Each feature corresponds to one reason for which a transaction may write a key. Code identity captures execution paths that always update the same state, and transaction arguments capture state selected by the input. Because an argument alone is shared with transactions that only read that state, RegionFence also combines arguments with code identity. All of these features are derived from the transaction itself, so every node can compute them before execution (Section~\ref{sec:implementation}).

The running example illustrates the need for combined features. $T_2$ and $T_3$ invoke the same router, so code identity does not distinguish swaps on P from swaps on other pools, and the address of P is also an argument of readers such as the price feed $T_1$. The combination of that address with the router is largely specific to swaps on P, so it identifies $T_2$ but not $T_1$.

The \emph{conflict profile} of a contended key $k$ stores a bounded number of writer features $\phi$ together with their estimates $\hat p_k(\phi)$. The estimate of a transaction is the maximum over its features. Because only features can identify a writer that has not yet accessed $k$, RegionFence uses them as the primary predictor and complements them with two kinds of direct evidence. Keys updated by read-modify-write are read before they are written, so an unfinished transaction that has read $k$ is a likely writer, with a probability given by the fraction of readers of $k$ that change it. An aborted transaction also provides exact evidence: its next execution treats earlier unfinished readers of the keys that invalidated it as predicted writers.

\textbf{Profile Maintenance.} RegionFence updates a profile as soon as evidence appears, because a key may become contended in the middle of a block. When a publication invalidates a read, the key joins $K_c$ immediately, and the features of the writer enter its profile. When a transaction first finishes, it records a trial for every tracked feature it carries on each contended key, and a hit if it changed the key. Each update is visible to the next fence, so a key that becomes contended within a block can identify its subsequent writers.

\textbf{Write and Abort Statistics.} RegionFence derives $\hat q_k$ and $\hat r_k$ from completed and committed transactions. Each completed execution adds its number of writes to each contended key to a histogram, from which $\hat q_k(m)$ is the fraction of executions with at least $m$ writes that have exactly $m$ writes. This estimate depends on how the code updates the key rather than on the caller. The estimate $\hat r_k$ is computed at commit by comparing the versions of aborted executions with the committed values, because an execution that read such a version may itself publish a transient value.

\subsection{Inter-Block Learning}
\label{sec:inter-block}

\textbf{Cross-Block State.} RegionFence retains its learned state across blocks, because contended keys largely recur. This state comprises $K_c$, the profiles, the write counts, and the change statistics. Retention introduces two problems: a key-feature pair initially has few samples, and old samples become misleading when the writers of a key change.

\textbf{Empirical Priors.} RegionFence addresses sparse samples with empirical Bayes shrinkage. Each observation of whether a finished transaction carrying $\phi$ changed $k$ is a Bernoulli trial, and with $h_{k,\phi}$ hits in $t_{k,\phi}$ trials the estimate is
\begin{equation}\label{eq:posterior}
\hat p_k(\phi) \;=\; \frac{h_{k,\phi} + s\,\mu_{k,\phi}}{t_{k,\phi} + s},
\end{equation}
the posterior mean under a prior with mean $\mu_{k,\phi}$ and fixed strength $s$. The prior mean is derived from the record of $\phi$ on other keys, shrunk toward the mean of its group, where a group is defined by the feature type and the number of transactions in the batch that carry the feature. Because the cost in Equation~\eqref{eq:rule} is linear in the probability, comparing the posterior mean with $p^\ast$ is the Bayes decision.

\textbf{Workload Shifts.} RegionFence adapts to changes in the writers of a key in two ways. It discards the evidence of a pair once a run of misses becomes statistically unlikely under the earlier hit rate, and it periodically decays the counts so that recent blocks dominate.

\subsection{Adaptive Execution Window}
\label{sec:window}

The window width controls the degree of concurrency across blocks. Because a single block offers limited parallelism, the blocks of a window $[b_{\min}, b_{\min}+W-1]$ execute concurrently while transactions commit in block order. A narrow window leaves cores idle while the next block waits, whereas a wide one causes thrashing because more transactions read stale values and re-execute. The optimal width varies with the workload, so RegionFence derives it from execution feedback.

After each block commits, RegionFence compares the benefit of one more block, $\Delta^{+}$, with the cost of the outermost block, $\Delta^{-}$. The term $\Delta^{+}$ is the worker time per worker that the window bound left idle during that block, weighted by the smoothed fraction of first executions in the outermost block that committed without re-execution. The term $\Delta^{-}$ is the time for which the commit frontier waited for re-executions of those transactions, plus the slowdown and additional re-execution caused by one more block, both predicted from recent measurements.

RegionFence changes the width by at most one block after each commit. With $N$ blocks to execute, the width is updated as
\begin{equation}\label{eq:window}
W \;\leftarrow\; \begin{cases} \min(W+1,\,N) & \text{if } \Delta^{+} > \Delta^{-},\\ \max(W-1,\,1) & \text{if } \Delta^{-} > \Delta^{+},\\ W & \text{otherwise,}\end{cases}
\end{equation}
and remains unchanged when no block is left to admit beyond the window. The width persists across blocks, and the single-step adjustment prevents overshooting after a burst.

\section{Implementation and Evaluation}
\label{sec:evaluation}

This section describes the implementation of RegionFence and evaluates it with respect to three questions.

\textbf{Q1: overall performance.} How does RegionFence perform on real workloads?

\textbf{Q2: conflict structure.} Which properties of conflicts determine the benefit of RegionFence?

\textbf{Q3: mechanism behavior.} Does each mechanism behave as designed?

\subsection{Implementation}
\label{sec:implementation}

We implement RegionFence as a package in go-ethereum v1.17.6, with a replay tool for recorded mainnet blocks and generated workloads.

\textbf{EVM Instantiation.} The general model of Section~\ref{sec:regions-fences} maps to the EVM as follows. Storage reads and balance reads are observations, and balance, nonce, and code are separate keys. A transfer into or out of a balance that the transaction does not read, such as the fee paid to the block beneficiary (coinbase), is a commutative update, and fees accumulate in a per-block layer. A check that a balance covers a transfer is a predicate read. While an earlier transaction of the same sender is uncommitted, the sender's nonce is taken from the transaction and checked at commit. The beneficiary's balance is the globally written key, and the slots of a contract form one object for contended key detection. Code identity consists of the called contract and the function selector. Transaction arguments are the aligned addresses in the calldata and in the EIP-2930 access list, and combinations pair an address or the sender with the called contract. The gas limit guarantees termination, and $\Gamma_b$ applies the non-transaction state changes of a block.

\textbf{Baselines.} Both baselines are separate executors in their pure forms, and they share only the EVM, field-level keys, and the deferred fee with RegionFence. Both execute one block at a time. The OCC baseline follows Block-STM~\cite{blockstm}: transactions execute on a multi-version memory, a read of an estimate aborts the execution until the writer re-executes, and validation tasks re-check read sets by version.

The PCC baseline uses strict two-phase locking with wound-wait. A transaction acquires shared locks for reads and exclusive locks for writes, and it holds them until it commits in block order. Locks are acquired at run time because the contracts of mainnet blocks declare no access sets, which precludes deterministic scheduling~\cite{calvin,bohm}. We choose wound-wait because deadlock detection scales poorly under contention~\cite{yu2014abyss,rebirth-retire}, whereas block order provides a natural priority that avoids deadlock.

\subsection{Experimental Setup}
\label{sec:setup}

\textbf{Platform.} All experiments run on a server with two 128-core AMD EPYC 9754 processors. A run on $P$ cores is pinned to $P$ physical cores of one socket, without SMT. Each configuration runs three times, and we report the median. Each trace starts with empty profiles and $\tau = 0.5$.

\textbf{Workloads.} We use nine mainnet traces from different periods of the past five years, replayed on their recorded pre-states, and two synthetic workloads. Database benchmarks such as TPC-C~\cite{tpcc} and SmallBank~\cite{smallbank} do not model contract calls that route through hot keys, so we design workloads for blockchains. ERC-20 generates token transfers under Zipf skew. DeFi-Mix combines price-feed reads of a hot pool, two-hop swaps through it, and unrelated transfers. A synthetic block holds 200 transactions, and a run executes 100 blocks. By default, DeFi-Mix directs 20\% of its transactions to the hot pool, half of them writing, with 16 storage accesses per transaction and the first hot read at 20\% of execution.

\textbf{Metrics.} We report the speedup of each system over its own single-core execution, computed as total time on one core divided by total time on $P$ cores, and the abort rate, the fraction of executions that abort. Aggregate results over the nine traces are based on total execution time.

\subsection{Performance on Mainnet Blocks}
\label{sec:real}

\begin{figure}[t]
\centering
\includegraphics[width=\columnwidth]{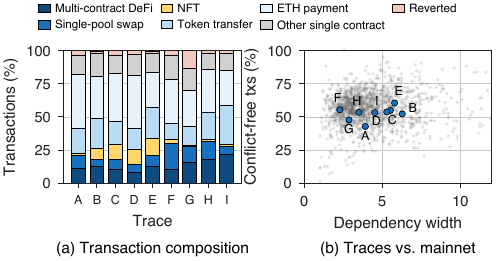}
\caption{Characteristics of the nine traces: (a)~transaction composition, and (b)~median per-block dependency width and conflict-free share, compared with about 1,400 sampled mainnet blocks.}
\label{fig:traces}
\Description{The first panel shows transaction composition on nine mainnet traces. The second places their median dependency width and conflict-free share against a background distribution of approximately fourteen hundred sampled mainnet blocks.}
\end{figure}

\begin{figure*}[t]
\centering
\includegraphics[width=\textwidth]{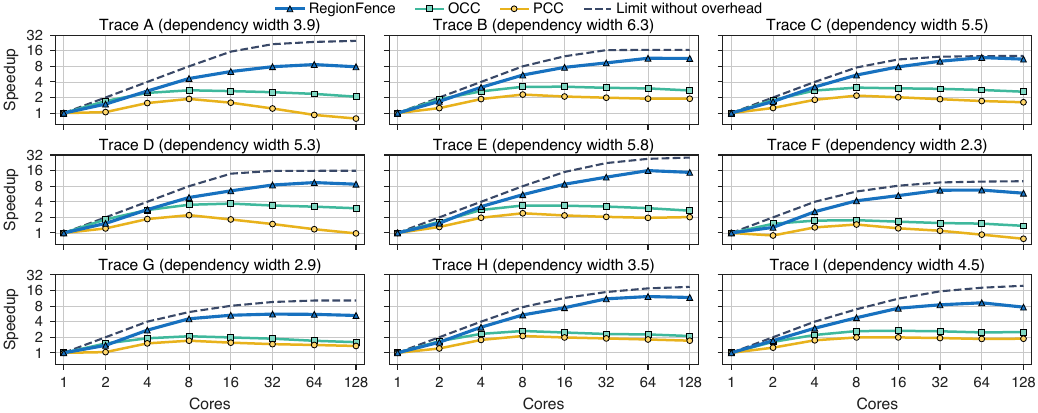}
\caption{Speedup on the nine traces, relative to each engine's own single-core execution. Dashed lines show the limit without overhead (Section~\ref{sec:real}).}
\label{fig:scaling}
\Description{Nine plots compare speedup against core count for RegionFence, OCC, and PCC on traces A through I. Each engine is normalized to its own single-core execution; dashed curves show the ideal region-level limit without overhead.}
\end{figure*}

We answer Q1 with the nine mainnet traces. RegionFence outperforms both baselines on all of them.

\textbf{Workload Diversity.} The nine traces span the range of mainnet workloads. Their transactions differ in composition, with multi-contract DeFi at 8\%--22\% and NFTs at 1\%--13\% (Figure~\ref{fig:traces}(a)). Their median blocks have dependency widths from 2.3 to 6.3 and 43\%--61\% conflict-free transactions, and they fall within the dense region of the distribution of sampled mainnet blocks (Figure~\ref{fig:traces}(b)). The results are therefore not specific to a particular application mix.

\textbf{Speedup.} RegionFence scales to 32--64 cores, whereas both baselines saturate at 8 cores (Figure~\ref{fig:scaling}). Over the nine traces, it reaches 9.2$\times$ at 64 cores and 8.3$\times$ at 128, whereas OCC peaks at 2.6$\times$ and PCC at 2.0$\times$. At 64 cores, RegionFence reaches 5.5--16.0$\times$ across traces, and its speedup is 2.9--5.4$\times$ that of OCC and 3.9--9.4$\times$ that of PCC on every trace.

\begin{figure}[t]
\centering
\includegraphics[width=\columnwidth]{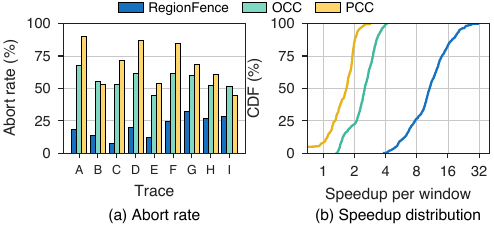}
\caption{Stability at 64 cores: (a)~abort rate on each trace, and (b)~CDF of speedup over sliding windows of 16 blocks.}
\label{fig:stability}
\Description{Two panels compare RegionFence, OCC, and PCC at 64 cores: abort rates on each trace, and cumulative distributions of speedup across sliding windows of sixteen consecutive blocks.}
\end{figure}

\textbf{Abort Rate.} RegionFence converts most aborts into short waits. At 64 cores, its abort rate is 7\%--32\% across traces, against 45\%--67\% for OCC and 44\%--90\% for PCC (Figure~\ref{fig:stability}(a)). OCC does not anticipate writers, so a reader that observes a stale version discards its entire execution. PCC also aborts transactions to resolve lock conflicts, which are frequent on hot keys. RegionFence instead delays a predicted reader until the writer's last write, which costs less than re-executing the transaction.

\textbf{Stability.} The performance gain is consistent across blocks. Over windows of 16 consecutive blocks, the median speedup is 10.6$\times$, and 95\% of the windows exceed the best window of OCC (Figure~\ref{fig:stability}(b)).

\textbf{Scalability Limit.} Further speedup is limited mainly by the parallelism of the blocks. The dashed lines show the \emph{limit without overhead}, an ideal region-level schedule of each trace with $P$ workers, serial transaction times, and the measured serial commit, relative to RegionFence's single core and capped at $P$. This limit reaches 92\%--100\% of its 128-core value at 64 cores, so speedup levels off near 64 cores on eight of the nine traces. RegionFence reaches 37\%--94\% of the limit at 64 cores, and the remaining gap reflects system overhead, mispredictions, and suspended transactions. At 128 cores, speedup falls by about 10\% in total because of overhead.

\subsection{Sensitivity to Conflict Structure}
\label{sec:synthetic}

\begin{figure*}[t]
\centering
\includegraphics[width=\textwidth]{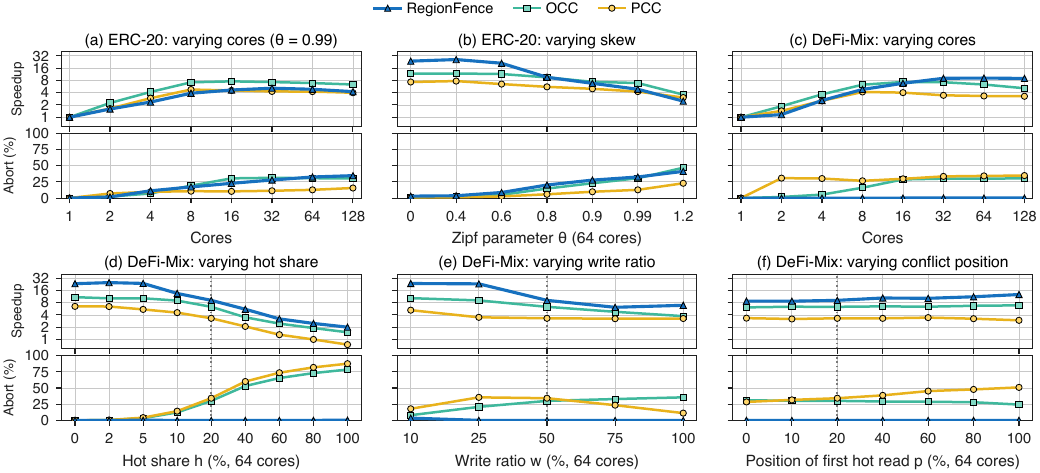}
\caption{Speedup (upper rows) and abort rate (lower rows) on synthetic workloads. Dotted lines mark the DeFi-Mix defaults.}
\label{fig:synthetic}
\Description{Six pairs of plots report speedup and abort rate for synthetic workloads as core count, contention skew, hot-transaction share, write ratio, or first-hot-read position varies. Dotted lines indicate default DeFi-Mix settings.}
\end{figure*}

Mainnet traces exhibit a mix of conflict types, so we answer Q2 with two synthetic workloads that isolate them. A conflict covers most of a transaction in ERC-20 and only part of it in DeFi-Mix, and we vary one property at a time.

\textbf{Contention Intensity.} The advantage of RegionFence under contention depends on how much of a transaction a conflict covers. On ERC-20 at 64 cores, RegionFence reaches 21.0--25.9$\times$ up to $\theta = 0.6$, about twice OCC, because most transactions are independent (Figure~\ref{fig:synthetic}(a)--(b)). The two systems perform equally at $\theta = 0.8$, and OCC outperforms RegionFence at higher skew, with 6.9$\times$ against 4.8$\times$ at $\theta = 0.99$ and no reduction in aborts. Under heavy skew, a region covers nearly an entire transfer and nearly every transfer joins a dependency chain, so waiting yields little parallelism, whereas fences and the window still incur overhead. On DeFi-Mix, a region covers only part of a transaction and the remaining operations stay parallel, so RegionFence stays 1.3--2.4$\times$ ahead of OCC as the hot share grows from 0\% to 100\% (Figure~\ref{fig:synthetic}(d)). Its abort rate remains near zero, whereas OCC and PCC reach 78\% and 88\%, because the writers of the hot key are predictable and fences wait before the conflict occurs.

\begin{figure}[t]
\centering
\includegraphics[width=\columnwidth]{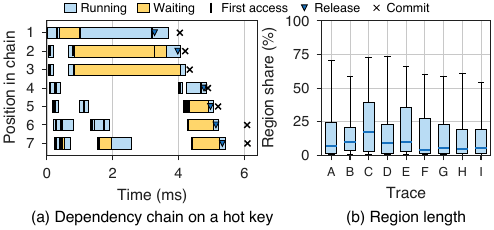}
\caption{Regions at 64 cores: (a)~timeline of the first seven transactions in the dependency chain on the most contended key, and (b)~region length relative to incarnation running time, for hot keys released early (boxes: quartiles; whiskers: 5th and 95th percentiles).}
\label{fig:region}
\Description{The first panel shows overlapping transactions on a hot-key dependency chain, with a release before the first commit. The second compares the region's fraction of transaction running time across nine traces, using quartiles and fifth-to-ninety-fifth percentile whiskers.}
\end{figure}

\begin{figure}[t]
\centering
\includegraphics[width=\columnwidth]{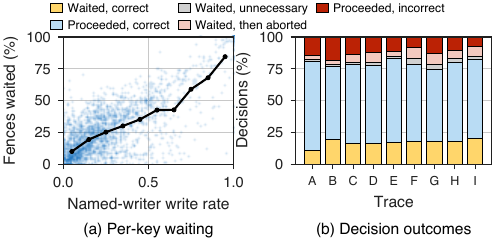}
\caption{Fence decisions at 64 cores: (a)~per-key share of fences that waited against the rate at which a named unfinished writer wrote the key (line: bin means), and (b)~decision outcomes on each trace.}
\label{fig:fence}
\Description{Two panels show fence decisions at 64 cores. The first relates the fraction of fences that wait on a key to the frequency with which a predicted writer actually writes it. The second breaks down decision outcomes for each of the nine traces.}
\end{figure}

\begin{figure*}[t]
\centering
\includegraphics[width=\textwidth]{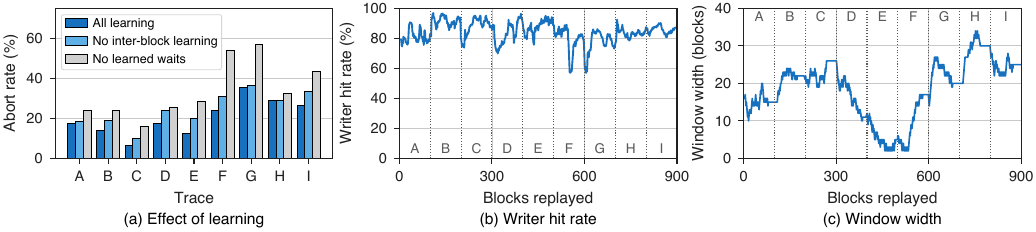}
\caption{Learning at 64 cores: (a)~abort rate with all learning, without inter-block learning, and without learned waits; (b)~share of waits whose named writer wrote the key (10-block moving average); (c)~window width. Panels (b) and (c) plot the traces in sequence.}
\label{fig:learned}
\Description{Three panels show the effect of learning on abort rate for each trace, the writer hit rate over the sequence of traces, and the adaptive window width over the same sequence. Writer hit rate uses a ten-block moving average.}
\end{figure*}

\textbf{Core Scaling.} The advantage of RegionFence grows with the number of cores. OCC scales better at low core counts, where conflicts are rare and the overhead of RegionFence dominates. As the number of cores increases, the re-execution cost of OCC grows, and RegionFence surpasses it at 32 cores and reaches 9.1$\times$ at 64 cores, against 6.3$\times$ for OCC and 3.3$\times$ for PCC (Figure~\ref{fig:synthetic}(c)).

\textbf{Write Ratio.} The advantage persists across write ratios, with RegionFence 1.3--2.5$\times$ ahead of OCC (Figure~\ref{fig:synthetic}(e)), because the write ratio changes the number of dependencies but not the fraction of a transaction that a region covers.

\textbf{Conflict Position.} A later conflict shortens the region and leaves more of a transaction independent. As the first hot read moves from the start to the end of a transaction, RegionFence rises from 8.7$\times$ to 12.7$\times$, whereas OCC and PCC remain nearly constant (Figure~\ref{fig:synthetic}(f)), because they control whole transactions regardless of where a conflict occurs. Only region-level control converts this independent work into parallelism.

The benefit of RegionFence thus depends on the fraction of a transaction that a conflict covers. It is largest when conflicts are partial, and it diminishes when nearly every transaction joins a dependency chain.

\subsection{Mechanism Analysis}
\label{sec:mechanisms}

To answer Q3, we replay the nine traces at 64 cores and record each decision.

\textbf{Regions.} Regions confine control to a small part of a transaction. Figure~\ref{fig:region}(a) shows the dependency chain on the most contended key. Although each transaction depends on its predecessor, the transactions overlap, and the first release precedes the first commit. Figure~\ref{fig:region}(b) shows that this holds in general: for hot keys released early, the median region covers 4\%--17\% of execution time. Dependent transactions thus overlap without re-execution, which neither OCC nor PCC achieves.

\textbf{Policy Mixture.} Fences apply different policies within one block (Figure~\ref{fig:fence}(a)). Across 2,461 keys, the share of fences that wait rises with the rate at which a predicted writer wrote the key, from about 10\% in the lowest bin to about 84\% in the highest. The cost rule produces this mixture without any per-key configuration.

\textbf{Decision Quality.} Most fence decisions are correct (Figure~\ref{fig:fence}(b)). In total, 62\% proceeded correctly and 18\% waited for a writer that wrote the key. Only 2.8\% waited unnecessarily, a superfluous constraint, whereas 10.6\% proceeded before a writer, a missing constraint, and 7.1\% aborted after a wait.

\textbf{Learning Effect.} Learning reduces aborts and execution time (Figure~\ref{fig:learned}(a)). Without inter-block learning, the mean abort rate increases from 20\% to 25\%, and without learned waits, to 34\%. Total execution time increases by 24\% and 22\%, respectively, so predictions directly affect performance.

\textbf{Prediction Accuracy.} Predictions recover after workload shifts (Figure~\ref{fig:learned}(b)). Overall, 83\% of waits target a writer that wrote the key. When this rate drops, as within trace F, it recovers within tens of blocks without an external signal, because evidence takes effect immediately and stale evidence is discarded.

\textbf{Window Adaptation.} The window follows the cost balance rather than the abort rate (Figure~\ref{fig:learned}(c)). Its width ranges from 2 to 34 blocks, with typical values of about 5 on trace E and about 30 on trace H. Traces G and C have the highest and lowest abort rates but the same median width of 22. The width is thus derived from execution feedback rather than fixed in advance.

\section{Related Work}
\label{sec:related-work}

\textbf{Fine-Grained Concurrency Control.} Prior work controls conflicts below the transaction level or exposes writes before commit. IC3 and PWV split transactions of known types into pieces that wait only for conflicting earlier pieces or expose writes early~\cite{ic3,pwv}, and Bamboo and Rebirth-Retire release a record after a transaction's last write to it~\cite{bamboo,rebirth-retire}. These protocols locate conflicts from known transaction types, annotations, or program analysis. Smart contracts are open and change continually, so RegionFence obtains this information during execution.

\textbf{Hybrid and Adaptive Concurrency Control.} Hybrid protocols select among policies for classes of data or work. MOCC selects by record temperature~\cite{mocc}, Sundial by conflict type~\cite{sundial}, CormCC by partition~\cite{cormcc}, Tebaldi by transaction type~\cite{tebaldi}, HDCC by transaction~\cite{hdcc}, and AOCC by operation~\cite{aocc}, and TxnSails adapts to the workload at run time~\cite{txnsails}. Each such choice covers a whole class. RegionFence instead applies one cost rule to each conflict, so different policies coexist within a block without predefined classes.

\textbf{Learned Concurrency Control.} Learned protocols derive the policy itself from the workload. Polyjuice trains a policy offline for stored procedures~\cite{polyjuice}, and AdaChain learns a system-wide architecture for permissioned blockchains~\cite{adachain}. Both learn what to do for a workload, which requires offline training or repeated trials. RegionFence learns conflicts rather than policies: it predicts writers online, and the cost rule makes each decision, so it follows workload shifts.

\textbf{Deterministic Concurrency Control.} Deterministic databases also execute in a predetermined serial order. Calvin, BOHM, and Caracal rely on read or write sets known in advance~\cite{calvin,bohm,caracal}, and Aria requires no such sets but reorders conflicting transactions across batches~\cite{aria}. RegionFence follows a preset order without access sets and without reordering.

\textbf{Orthogonal Blockchain Techniques.} RegionFence determines when a conflicting read occurs, and several techniques improve other stages of block processing. Pre-execution~\cite{forerunner,seer} and hardware acceleration~\cite{mtpu} shorten each execution, and GPU-based state commitment~\cite{mpt-gpu} accelerates the stage after execution. Block-building systems~\cite{fafo,anthemius} reduce conflicts before a block forms, and semantic approaches~\cite{rapidlane,octopus,crystality} remove specific classes of conflicts, so RegionFence handles the remaining ones. Operation-level re-execution~\cite{parallelevm} could reduce the cost of its aborts, and block-level access lists~\cite{eip7928} could seed conflict profiles.

\section{Conclusion}
\label{sec:conclusion}

This paper presents RegionFence, an adaptive concurrency control mechanism for blockchain parallel transaction execution. Regions confine control to the span of a transaction that a dependency constrains, and fences decide at run time, based on measured costs, whether each read proceeds or waits for the writer's last write. Intra-block and inter-block online learning predicts writers and determines the number of concurrent blocks from execution itself. Because certification is separate from scheduling, any mixture of policies within a block produces the serial result. Our evaluation on real Ethereum mainnet blocks from different periods of the past five years shows that RegionFence achieves an overall speedup of 9.2$\times$ over single-core execution on 64 cores and outperforms OCC by 2.9--5.4$\times$ and PCC by 3.9--9.4$\times$ in speedup across these periods.

\begin{acks}
This work was supported by the National Natural Science Foundation of China No. 62602642 and No. U22B2032.
\end{acks}

\clearpage
\bibliographystyle{ACM-Reference-Format-citeorder}
\bibliography{references}
\end{document}